\documentclass[aps,prx,superscriptaddress,a4paper,notitlepage,twocolumn]{revtex4-2}
\pdfoutput=1
\usepackage{ifpdf}

\usepackage{graphicx}
\usepackage{dcolumn}
\usepackage{bm}
\usepackage{amsmath}
 
\graphicspath{{/}}
\usepackage[dvipsnames]{xcolor}
\usepackage{cancel}
\usepackage{upgreek}
\usepackage{braket}
\usepackage{eurosym}
\usepackage[colorlinks=true,citecolor=Periwinkle,urlcolor=Periwinkle,linkcolor=BlueViolet]{hyperref}
\usepackage{tabularx}

\makeatletter
\renewcommand\tableofcontents{%
    \@starttoc{toc}%
}
\makeatother

\begin{document}
\title{Magnetic excitons in a suspended 2D antiferromagnetic membrane}
\author{Joanna L.P. Wolff}
\affiliation{Universit\'e de Strasbourg, CNRS, Institut de Physique et Chimie des Mat\'eriaux de Strasbourg (IPCMS), UMR 7504, F-67000 Strasbourg, France}
\author{Loïc Moczko}
\affiliation{Universit\'e de Strasbourg, CNRS, Institut de Physique et Chimie des Mat\'eriaux de Strasbourg (IPCMS), UMR 7504, F-67000 Strasbourg, France}
\author{Jérémy Thoraval}
\affiliation{Universit\'e de Strasbourg, CNRS, Institut de Physique et Chimie des Mat\'eriaux de Strasbourg (IPCMS), UMR 7504, F-67000 Strasbourg, France}
\author{Michelangelo Romeo}
\affiliation{Universit\'e de Strasbourg, CNRS, Institut de Physique et Chimie des Mat\'eriaux de Strasbourg (IPCMS), UMR 7504, F-67000 Strasbourg, France}
\author{Benjamin Bacq-Labreuil}
\affiliation{Universit\'e de Strasbourg, CNRS, Institut de Physique et Chimie des Mat\'eriaux de Strasbourg (IPCMS), UMR 7504, F-67000 Strasbourg, France}
\affiliation{Institut Quantique, D\'epartement de physique $\&$ RQMP, Universit\'e de Sherbrooke, Qu\'ebec J1K2R1, Canada}
\author{Stéphane Berciaud}
\affiliation{Universit\'e de Strasbourg, CNRS, Institut de Physique et Chimie des Mat\'eriaux de Strasbourg (IPCMS), UMR 7504, F-67000 Strasbourg, France}
\author{Arnaud Gloppe}
\email[]{arnaud.gloppe@cnrs.fr}
\affiliation{Universit\'e de Strasbourg, CNRS, Institut de Physique et Chimie des Mat\'eriaux de Strasbourg (IPCMS), UMR 7504, F-67000 Strasbourg, France}

\begin{abstract} 
Layered magnetic and strongly correlated materials present a rich platform for condensed matter physics with intrinsic properties intertwined by magnetism and low-dimensionality. A suspended light-emitting 2D antiferromagnetic membrane forms a highly controllable hybrid system in which the interplay between spin ordering, optical and mechanical degrees of freedom can be uniquely explored. NiPS$_3$ hosts excitons responsible for a puzzling luminescence down to a few atomic layers, linked to its zigzag antiferromagnetic order. The nature of these excitons remains unclear. Here we report on the magnetic excitons of a suspended few-layer NiPS$_3$ membrane. We reveal nematic zigzag states and study the optical transitions induced by the magnon-mediated motion of the excitation in the magnetic lattice, resulting in magnon-dressed excitons. We observe a strain tuning of these emission lines, dependent on their microscopic origin, with rates that sign a strong localization of the magnetic excitons, fading with the number of magnon-mediated hops.
\end{abstract}

\maketitle
Van der Waals magnets offer a large variety of magnetic orders (as intra- and interlayer ferro- and antiferromagnetism) and configuration that can be described along Ising, Heisenberg or XY  models, from textbook phenomenology to more exotic spin states~\cite{Augustin2021,Dupont2021}. Beyond providing a formidable platform for advanced condensed matter, these structures enable the development of ultimately-thin devices for applications in nano-spintronics and photonics. Their magnetic degrees of freedom can be integrated in functionalized van der Waals heterostructures~\cite{Huang2020}, made of the stacking of other layered materials that can be, for example, direct bandgap semiconductor as some transition metal dichalcogenides monolayers, exhibiting an optical response strongly dependent on their magnetic state~\cite{Zhong2017,Ciorciaro2020}. 

By tuning lattice parameters and orbital overlaps, macroscopically-controlled microscopic deformations are instrumental in exploring the properties of atomically-thin materials, as the Coulomb-dominated exciton physics of transition metal dichalcogenides~\cite{Mak2018a,Lopez2022, Wu2025}, 
as well as the behavior of magnetic van der Waals materials, dictated by the interplay of anisotropies and exchange interactions
~\cite{Jiang2020,Siskins2020, Houmes2023,Lyons2023,Hwangbo2024}. 
Hybrid optomechanical nanosystems, formed by a nanomechanical resonator coupled to spin degrees of freedom interacting with light, have been developed over the last decades towards the detection of subtle magnetic interactions~\cite{Rugar2004}, the potential transfer of non-classical states to the mechanical modes and the fine tuning of their quantum properties~\cite{Arcizet2011,Yeo2014,Barfuss2015}. 

\begin{figure*}
\includegraphics[scale=1]{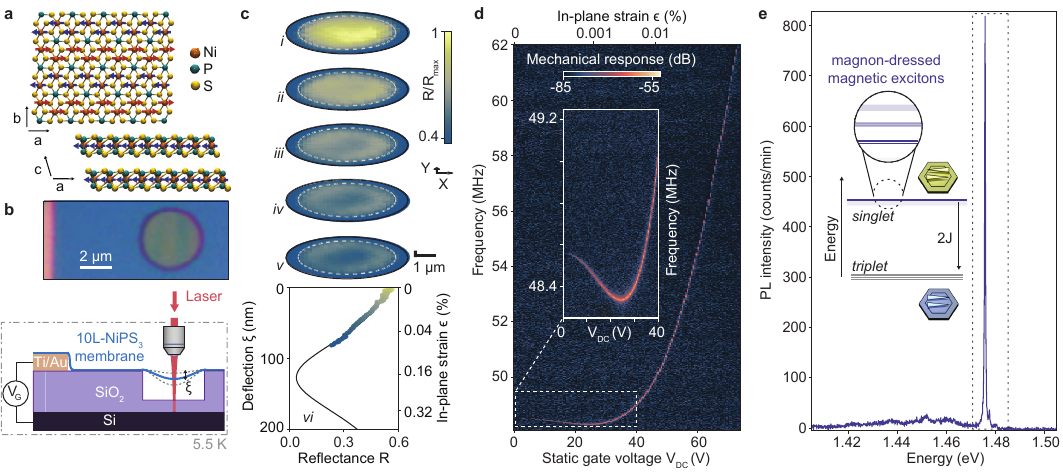}
\caption{\label{fig:fig1}\textbf{A light-emitting drum-like antiferromagnetic membrane.} \textbf{a,} The magnetic structure of NiPS$_3$ (top and side views) displays a zigzag antiferromagnetic order supported by the Ni$^{2+}$ ions in a preferential direction determined by the monoclinic stacking along \textit{c}~\cite{Ouvrard1985}. 
\textbf{b,} Optical picture of a 10-layer NiPS$_3$ suspended membrane on a Si/SiO$_2$ pre-patterned substrate (see Methods). The membrane is addressed by a focused laser beam in vacuum at 5.5\,K and contacted by a gold electrode to apply a gate voltage $V_\mathrm{G} = V_\mathrm{DC} + V_\mathrm{AC} \cos \Omega t$. 
 \textbf{c,} Spatially-resolved static reflectance of the membrane, with its boundaries marked by white dashes, measured as a function of the applied static gate voltage (\textit{i}-\textit{v}: $V_\mathrm{DC} =$ 0, 54, 64, 70 and 74\,V). The system reflectance is modeled as a function of the membrane deflection, plotted as a solid line in \textit{vi}, with the relative reflectance measured at the center of the membrane appearing as color dots as a function of the static gate voltage from  0\,V (yellow) to 75\,V (blue). The integrated static in-plane strain $\epsilon$ can be then estimated (see SI). \textbf{d,} Mechanical response of the membrane fundamental mode to an electrostatic drive upon increasing static voltage ($V_\mathrm{AC} = 10\,$mV). The inset is a zoom on the low-voltage region. 
\textbf{e,} Photoluminescence spectrum from the center of the 10-layer NiPS$_3$ membrane at $V_\mathrm{DC}=0$\,V. The dashed rectangle corresponds to the region of interest with emission peaks exhibiting sub-meV linewidths, originating from a triplet-to-singlet magnetic excitation with $J$ Hund's exchange energy ($2J \sim 1.47$\,eV). The two arrows within hexagonal tiles picture the particles spin in the two relevant effective orbitals centered on a Ni site, forming the triplet ground state or the singlet superposition, in blue and yellow respectively.} 
\end{figure*}

NiPS$_3$, a transition metal phosphorus trichalcogenide, is a van der Waals semiconductor presenting an intralayer zigzag antiferromagnetic order and an interlayer ferromagnetic alignment along the \textit{c}-direction of its monoclinic stacking, below 155\,K, in the bulk (Fig.~\ref{fig:fig1}\textbf{a}). Ni$^{2+}$ ions form a hexagonal lattice in the $ab$ plane, their magnetic moments pointing along the $a$-direction with a small out-of-plane component, modeled as an approximate XXZ system~\cite{Balkanski1987, Kim2019a, Wildes2022}. 	

It features a very narrow luminescence line at 1.475\,eV in its antiferromagnetic phase with typical linewidths below 500\,$\mu$eV~\cite{Kang2020,Jana2023,Lebedev2025}, polarized orthogonally to the zigzag direction in the bulk~\cite{Wang2021,Hwangbo2021}. 	
Broken by the monoclinic stacking which forces the zigzag direction, the three-fold hexagonal symmetry restores towards the monolayer limit, turning few-layer NiPS$_3$ into a promising system in which the magnetic moments can align along a discrete set of directions or nematic states~\cite{Hwangbo2024,Sun2024,Tan2024}. 

While seemingly related to the zigzag magnetic order, the origin of this photoluminescence is still under debate~\cite{Kang2020,Jana2023,He2024}, triggering different theoretical approaches~\cite{Klaproth2023,Scheie2023, Hamad2024}. Along with the main emission peak, stand a few others that have been largely overlooked, sharing similar features as a narrow linewidth and peculiar polarization properties. Thus, they appear as essential to understand the microscopic mechanisms at play in this correlated antiferromagnet.

Here, exerting tunable electrostatically-induced strain to a suspended membrane of few-layer NiPS$_3$, we elucidate the origin of its different transitions, observing the spatially-resolved evolution of the emission energies and polarizations. We reveal the coexistence of two nematic families corresponding to two zigzag directions. While the main magnetic excitons appear to be very localized, at the level of the lattice parameter, the enhanced strain sensitivity of higher order magnetic excitons signs their delocalization, 
originating from the displacement of the excitation in the magnetic environment where it is dressed by magnons through local spin flips in the magnetic zigzag order.  

Our system is depicted in Fig.~\ref{fig:fig1}\textbf{b}, made from a membrane of few-layer NiPS$_3$, suspended on a $5\,\mu$m-diameter pre-patterned hole in a Si/SiO$_2$ substrate and contacted to Ti/Au electrodes (see Methods). A 633-nm HeNe laser is focused on the membrane, allowing concomitantly the detection of its nanomechanical properties by optical interferometry and optical spectroscopy. 
The membrane is constituted by ten NiPS$_3$ layers ($\sim$ 9\,nm-thin), to maximize its mechanical susceptibility and weaken the stacking-induced anisotropy, while preserving a clear photoluminescence spectrum, which otherwise degrades as the monolayer limit is approached~(see down to a bilayer in SI). The measurements are done in vacuum at 5.5\,K, well below NiPS$_3$ Néel temperature.
\vspace{-1mm}
\section*{Results}
\subsection*{A light-emitting antiferromagnetic membrane}
Suspending the few-layer flake allows for a fine electrostatic control of its deformation. A static gate voltage $V_\mathrm{DC}$ between the gold electrode and the Si substrate induces an electrostatic in-plane strain,  resulting from the vertical deflection $\xi$ of the membrane towards the cavity bottom, see Fig.~\ref{fig:fig1}\textbf{b}. The length of the optical cavity formed by the membrane and the substrate depends on the vertical deflection, modifying the membrane reflectance $R$. Figure~\ref{fig:fig1}\textbf{c} shows the spatially-resolved membrane reflectance at 633\,nm, for different static gate voltages, measured on a photodiode. By comparing the observed change in reflectance to a multi-interface model based on Fresnel equations (see SI), we estimate the induced deflection at the center of the membrane to $\xi_c\sim80$\,nm at $V_\mathrm{DC}$~=~75\,V, resulting in an integrated $\epsilon \sim 0.07$\% 
in-plane strain.
The membrane vibrations are electrostatically-driven through a gate voltage $V_\mathrm{G} = V_\mathrm{DC} + V_\mathrm{AC}\cos\Omega t$ (with $\Omega/2\pi$ the drive frequency) and read by optical interferometry (mechanical quality factors $Q \sim 10^3 - 10^4$) ~\cite{Aspelmeyer2014,Steeneken2021}. Figure~\ref{fig:fig1}\textbf{d} presents the evolution of the mechanical response of the membrane as a function of the applied static voltage, acquired through a vector network analyzer. The strain modifies the membrane fundamental mode eigenfrequency, originally at 48.6\,MHz, inducing first a softening to compensate for built-in tension before following a continuous parabola as a function of $V_\mathrm{DC}$. 
This behavior is typical of an ideal capacitively-coupled membrane, making the strain traceable in our experiment (see SI).  

A photoluminescence (PL) spectrum of the membrane, shown in Fig.~\ref{fig:fig1}\textbf{e}, reveals the main narrow peak at 1.475\,eV accompanied by a broad structure at lower energy attributed to phonon sidebands~\cite{Wang2021} and a subtle structure at higher energy spreading on a few meV that will be discussed in the following. 
The main magnetic exciton has been simplistically described as an electron transiting to Ni ion $d$ orbitals from S $p$ orbitals, leaving a hole there~\cite{Wang2021}. The S ligands experience a residual magnetization depending on their position relatively to two Ni zigzag antiferromagnetically coupled chains, resulting in an anisotropic linear emission in the bulk and in thin films~\cite{Dirnberger2022}. 
Ab initio calculations, combined with a low-energy effective $t-J$ model, predict the exciton to be a triplet-to-singlet excitation involving two wavefunctions centered on Ni sites and spreading on adjacent P and S atoms~\cite{Hamad2024}. In this scenario, the excitation can move from site to site, inducing spin flips along the way. They appear as defects in the local magnetic configuration which witness the dressing of the exciton by magnons. This magnon-assisted hopping is analogous to the hole motion in high-temperature superconductors~\cite{Kane1989,Martinez1991}.
\subsection*{Optically-detected nematic families in the few-layer NiPS$_3$ membrane} 
\begin{figure*}
\includegraphics[scale=.95]{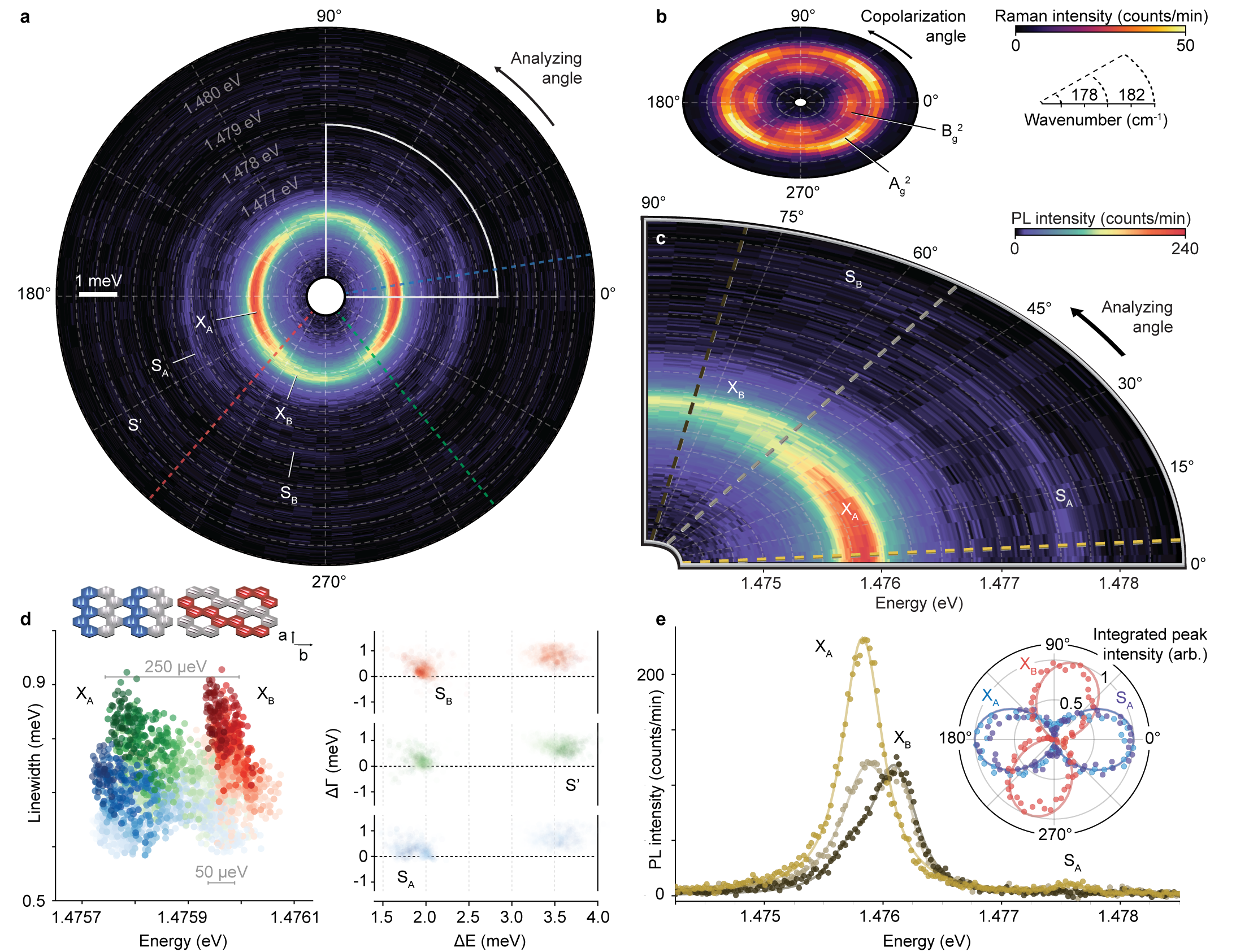}
\caption{\label{fig:fig3}\textbf{Polarization of the light emitted by a few-layer NiPS$_3$ membrane revealing nematic families.} \textbf{a}, Photoluminescence spectra from the center of the membrane for analyzing angles between 0 and 360$^\circ$ revealing transitions from two distinct nematic families (labeled with subscripts A and B). \textbf{b}, Polarized Raman spectroscopy at the center of a few-layer NiPS$_3$ membrane involving the phonon modes A$_g^2$ and B$_g^2$ used as the angle reference. 
\textbf{c}, Zoom on the PL spectra from 0$^\circ$ to 90$^\circ$. 
The color dashed lines indicate the location of the spectra shown in \textbf{e}. \textbf{d}, Extracted energy and Voigt full width at half maximum of the main peaks from the membrane and its immediate surroundings for three different analyzing polarization 
marked in red, green and blue dotted lines in \textbf{a}. The top sketches represent the magnetic ground state of nematic families A and B. The transparency scales on the peak relative intensity, the color scales with the DC reflectance to highlight the suspended region (see SI).
For each analyzing polarization, the S and S' peaks properties are reported as a function of the relative energy and linewidth of the main peak. \textbf{e}, Photoluminescence spectra for analyzer angles of 4$^\circ$ (gold), 54$^\circ$ (silver) and 79$^\circ$ (bronze). Solid lines are fit to Voigt profiles (see SI). Inset: normalized integrated intensity in the range $1.4758\pm 1\times10^{-4}$\,eV for X$_\mathrm{A}$ (light blue dots), $1.4761\pm 1\times10^{-4}$\,eV for X$_\mathrm{B}$ (red dots), and $1.4776\pm 3\times10^{-4}$\,eV for S$_\mathrm{A}$ (purple dots) as a function of the analyzing angle $\theta_A$. The solid lines are fit to $\cos^2(\theta_A - \theta_{\mathrm{X_i}})$, leading to the relative polarization angles $\theta_{\mathrm{X_A}} = 0^\circ = \theta_{\mathrm{S_A}}$ and $\theta_{\mathrm{X_B}} = 72^\circ$. 
}
\end{figure*}
We investigate the polarization of the photons emitted by the few-layer membrane to clarify the nature of the different transitions and later study their specific evolution with the electrostatically-induced strain. 
The incident light is linearly polarized. It has been previously reported that NiPS$_3$ emission peak PL does not show any dependence with the incoming polarizing angle~\cite{Wang2021}. We collect photoluminescence spectra for various angles of an analyzer placed before the optical spectrometer (see Methods).

Two phonon modes, noted B$_g^2$ and A$_g^2$ at 179\,cm$^{-1}$ and 181\,cm$^{-1}$, sensitive to the magnetic order~\cite{Kim2019a,Sun2024, Chen2024}, are observed by polarization-resolved Raman spectroscopy in Fig.~\ref{fig:fig3}\textbf{b} (see Methods and SI). The 4-fold symmetry displayed by the Raman response of B$_g^2$, corresponding to a displacement of Ni atoms by respect to the main zigzag axis~\cite{Hashemi2017,Kim2019a}, is used as the angle reference in the PL polarization polar plots.

\begin{figure*}
\includegraphics[scale=1]{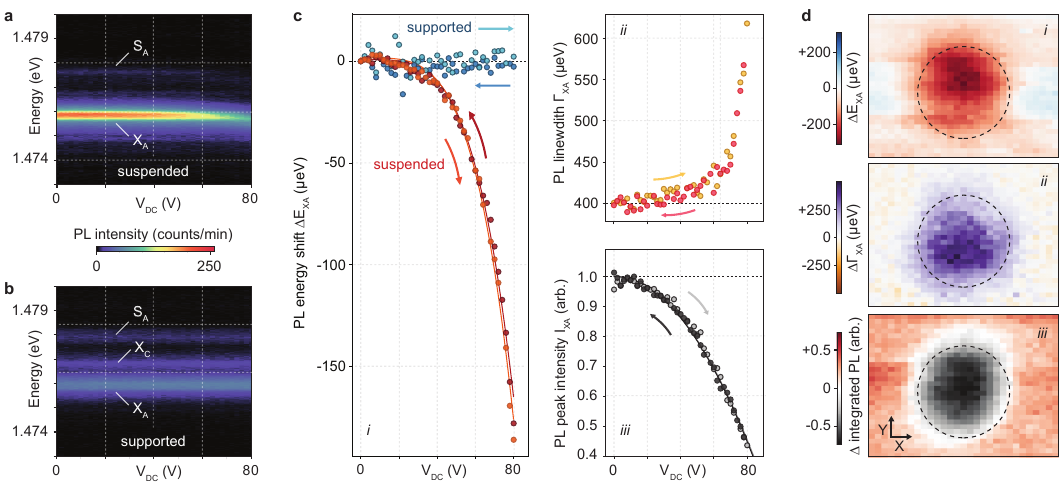} 
\caption{\label{fig:fig2}\textbf{Strain-tuned photoluminescence of the main magnetic exicton in a few-layer NiPS$_3$ membrane.} \textbf{a}, Photoluminescence spectra zoomed on the main emission peak as a function of the static gate voltages $V_\mathrm{DC}$ from 0 to 80\,V, with an analyzing polarization favoring X$_\mathrm{A}$, at the center of the membrane. \textbf{b}, Photoluminescence spectra acquired on a nearby supported area in the same conditions. \textbf{c}, Comparison of the extracted photoluminescence peak energy shifts $\mathrm{E_{XA}}-\mathrm{E_{XA}}(0\,\mathrm{V})$ as a function of increasing and decreasing static gate voltages at the center of the membrane (red) and at nearby supported location (blue) \textit{(i)}. The solid lines are fits in $a_V V^4_\mathrm{DC}$, where $a_V =-4.3$\,peV/V$^4$ ($-4.0$\,peV/V$^4$) for increasing (decreasing) voltages, corresponding to an energy shift of meV/\% of radial strain. The insets \textit{(ii)} and \textit{(iii)} show respectively the variation of linewidth and normalized peak intensity extracted from the suspended membrane photoluminescence. The intensity follows $V_\mathrm{DC}^2$ as the vertical deflection of the membrane. \textbf{d,} Spatially-resolved comparison of the photoluminescence spectra in the vicinity of the drum, for a static gate voltage of 75 and 0\,V: \textit{(i)} the energy, \textit{(ii)} the linewidth and \textit{(iii)} the integrated photoluminescence. The dashed lines represent the membrane contour determined from the DC optical reflectance at 633\,nm measured simultaneously on a photodiode.}
\end{figure*}

Figure~\ref{fig:fig3}\textbf{a} presents such PL spectra acquired at the center of the drum. 
In addition to the main peak (X$_\mathrm{A}$) at 1.4758\,eV, a distinct peak (X$_\mathrm{B}$) appears very close by, about 0.3\,meV higher. Its intensity is about half of the main peak, its linewidth is comparable. Both transitions lead to a linearly polarized emission, with an angle difference of $\sim72^\circ$, hinting for another zigzag domain allowed by the reduced number of stacked layers. The slight lift of degeneracy and the angle deviation from an ideal 3-fold hexagonal symmetry may be due to a reminiscence of the monoclinic stacking, as well as nanoscopic defects and corrugations~\cite{Lopez2023}. A dimmer peak (S$_\mathrm{A}$) appears 1.8\,meV above X$_\mathrm{A}$, with which it shares the same polarization properties. This suggests a dressing of the main excitation by magnon-assisted hopping between neighboring antiferromagnetic zigzag chains. 
The existence of such a zone-center magnon, around 1.5\,meV, has been hinted by Raman and inelastic neutron scattering measurements on supported few-layers~\cite{Scheie2023, Jana2023, Na2024}. 
The magnon-mediated hopping of the main excitation was foreseen from first principle calculations, which reveal that the dominant hopping term in a two-band Hubbard model corresponds to the third nearest neighbor, coupling different zigzag chains~\cite{Scheie2023,Hamad2024}.

We chart the discernible transitions all over the membrane and report on Fig.~\ref{fig:fig3}\textbf{d} the energy and linewidth of PL peaks acquired on the drum and its immediate vicinity ($\sim 9\times9\,\mu$m$^2$ area), 
 extracted from spectra taken with three different analyzing polarization to cover for the expected hexagonal 3-fold symmetry. We are able to distinguish the two nematic families. The equivalent secondary peak S$_\mathrm{B}$ for X$_\mathrm{B}$, appearing faintly in Fig.~\ref{fig:fig3}\textbf{a}, lies at about 1.9\,meV from X$_\mathrm{B}$ with the same polarization properties.   
Higher in energy, at about 3.5\,meV, stands another tenuous peak S' with a larger linewidth and no clear polarization dependence (at $V_\mathrm{DC} = 0$). 
Corresponding to about twice the magnon energy, we assume this feature to be a mixture of higher order transitions involving two successive hops of the excitation over two chains, that can branch in a variety of directions, resulting in a larger scattering of its energy and apparent linewidth. Peaks with similar features in energy and linewidth as S and S' have been theoretically predicted by taking into account exchange and hoppings terms between nearest neighbors~\cite{Hamad2024}. 
The energy of the main exciton peaks X$_\mathrm{A}$ and X$_\mathrm{B}$ over the drums typically spread over 50$\,\mu$eV, which is constricted enough to experimentally discriminate these two nematic states using the proper analyzing angle. 
We note the existence of a peak X$_\mathrm{C}$ appearing mainly on the surrounding supported part (visible in Fig.~\ref{fig:fig2}\textbf{b}), 1\,meV higher from the main peak, that may be related to the third zigzag direction. 
We observe a similar relative polarization pattern on a likewise drum for which X$_\mathrm{C}$ appears instead of X$_\mathrm{B}$ (see SI). 
\subsection*{Strain tuning of the main magnetic exciton}
\begin{figure*}
\includegraphics[scale=1]{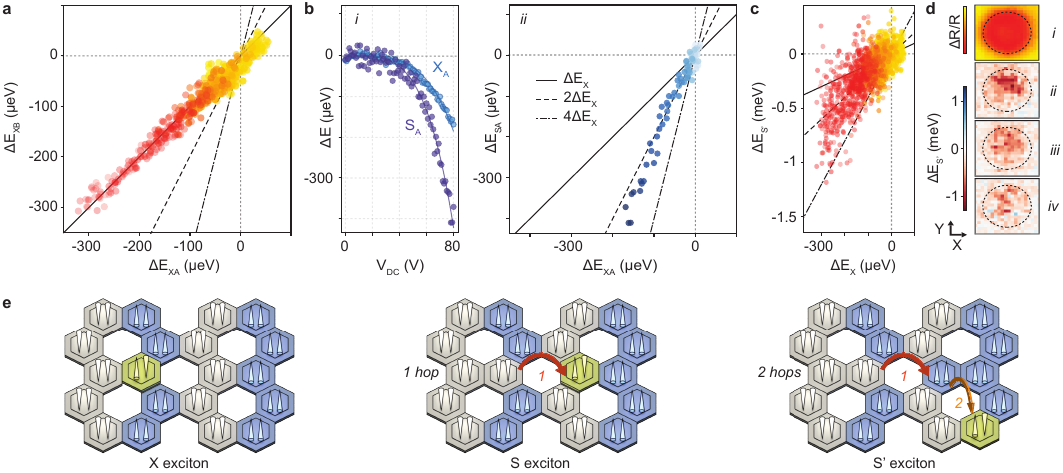}
\caption{\label{fig:fig4}\textbf{Strain tuning and nature of the different magnetic excitons.} \textbf{a}, Spatial correlation between the energy shifts observed on X$_\mathrm{B}$ and the energy shifts on X$_\mathrm{A}$ at 75\,V (0\,V being the reference) over the membrane and its surroundings. The transparency scales on the peak relative intensity, the color scales with the DC reflectance (see \textbf{d} and SI). Lines with slopes of 1, 2 and 4 are indicated as a guide for the eyes. \textbf{b},
 \textit{(i)} Energy shift for S$_\mathrm{A}$ (purple), and corresponding X$_\mathrm{A}$ (light blue), with 0\,V as the reference. The solid lines are fits in $a_V V^4_\mathrm{DC}$, where $a_V^\mathrm{SA} =-10.3$\,peV/V$^4$ and $a_V^\mathrm{XA} = -4.5$\,peV/V$^4$.  \textit{(ii)} Correlations between the shift in X$_\mathrm{A}$ and S$_\mathrm{A}$. 
\textbf{c}, Spatial correlation between the energy shifts at 75\,V (0\,V being the reference) of S' and X. The transparency scales on S' peak relative intensity, the color scales with the DC reflectance.
\textbf{d,} \textit{(i)} DC reflectance used to map the spatial correlation in \textbf{a} and \textbf{c}, highlighting in red and in yellow the suspended and supported parts, respectively. \textit{(ii-iv)} Spatially-resolved energy shifts on S' analyzed with the three analyzing polarizations shown in Fig.~\ref{fig:fig3} (red, green, blue, favoring $X_\mathrm{B}$ for the former and $X_\mathrm{A}$ for the latter).
\textbf{e}, Sketches of the magnetic configuration before the recombination of the X exciton (bare exciton), of the S exciton after one hop of the excitation (exciton dressed by one magnon) and of the S' exciton after two hops of the excitation (exciton dressed by two magnons), respectively.}
\end{figure*}
We examine the nature of these excitons through local strain. We focus first on the main emission peak (X$_\mathrm{A}$) and demonstrate that its emission energy is weakly downshifted by the electrostatically-induced strain.
 
Figure~\ref{fig:fig2}\textbf{a} shows photoluminescence spectra acquired from the center of a NiPS$_3$ nanodrum, with an analyzing polarization favoring X$_\mathrm{A}$, at different static gate voltages $V_\mathrm{DC}$. The spectra are adjusted with a Voigt profile and their fitting parameters extracted in Fig.~\ref{fig:fig2}\textbf{c}. 
The peak energy decreases by 0.18\,meV from 0 to 80\,V, corresponding to about half of the peak initial apparent linewidth (0.4\,meV). This red shift follows $V_\mathrm{DC}^4$ as the integrated in-plane radial strain $\epsilon$ (see SI). It translates into an energy shift with the radial strain in the $-1$\,meV/\% range. 
The peak PL intensity is halved and follows $V_\mathrm{DC}^2$, coinciding with the membrane deflection $\xi_c$ as a result of the interferences within the optical cavity foreseen in Fig.~\ref{fig:fig1}\textbf{c} (see SI). 

We acquire the spectra for decreasing gate voltages to check for hysteretic behaviors. The extracted intensity, energy and linewidth in both directions overlap, highlighting the reversibility of the strain tuning in this deflection regime. Predominant contributions from doping effects are ruled out by carrying out the same measurements on a non-suspended part of the sample, at the same distance from the electrode as the center of the drum (Fig.~\ref{fig:fig2}\textbf{b}). At this location, there is no significant change in the photoluminescence spectra with the applied gate voltage (Fig.~\ref{fig:fig2}\textbf{c}).

We collect spatially-resolved photoluminescence spectra from the drum and its surroundings. In Fig.~\ref{fig:fig2}\textbf{d}, we compare the integrated photoluminescence intensity, the energy and the linewidth of the main emission peak X$_\mathrm{A}$ at a static gate voltage of 75\,V and 0\,V, used as a reference. The major variations are located within the membrane region, attesting that the observed features are due to the electrostatically-induced strain on the membrane. 
The peak linewidth is slowly increasing until 60\,V, where it rises sharply. This broadening arises from the strain gradient over the optically probed area of the deflected membrane.
On some supported parts, maybe due to doping, the energy shifts slightly towards higher values (in the $+50\,\mu$eV range at 75\,V).

\subsection*{Strain sensitivity and localization of the excitons} 
The energy variation map of X$_\mathrm{A}$ deviates from the expected circular symmetry of the radial strain. 
We observe a similar feature for X$_\mathrm{B}$ (see SI). Tracking a possible axial dependency of the strain sensitivity with the zigzag direction, we scrutinize the effect on X$_\mathrm{B}$. We compare position by position its shift in energy from 0 to 75\,V to the shift on X$_\mathrm{A}$, over the membrane and its surroundings, in Fig.~\ref{fig:fig4}\textbf{a}. These correlations suggest that the microscopic changes induced by the strain on the excitonic environment are not sensitive to the zigzag direction and thus that the excitons are very localized.
We also perform polarization-dependent PL measurements as the strain increases to monitor possible domain switches (see SI). We do not discern any significant change in the photoluminescence polarization distribution as the strain increases, in our conditions. The strain sensitivity of the main exciton peaks X$\mathrm{_A}$ and X$\mathrm{_B}$ are equivalent. 

We report in Fig.~\ref{fig:fig4}\textbf{b} the evolution of the energy for S$_\mathrm{A}$ with the static gate voltage. The maximum observed shift is twice the shift observed on the main peak.  
This enhanced sensitivity suggests that the S magnetic excitons are less localized than X, in line with a magnon-mediated excitation, depicted in Fig.~\ref{fig:fig4}\textbf{e}. Its PL peak intensity follows the membrane deflection (see SI). As for X, no intrinsic change in the radiative yield from the evolution of the excitonic environment is noticeable for S neither. This specific measurement is done on another drum. Remarkably, the strain sensitivities between the two distinct drums, both for X$\mathrm{_A}$ ($-4.2\pm0.3\,$peV/V$^4$) and for S$\mathrm{_A}$ ($-9\pm1.5\,$peV/V$^4$), are very consistent (see SI). 

Finally, we consider the strain sensitivity of the peak S' in Fig.~\ref{fig:fig4}\textbf{c} by plotting the spatial correlation of its energy shift with X. The maximum shift observed reaches $-1.2$\,meV, corresponding to about two times the sensitivity of S excitons and four times the sensitivity of X excitons. Signing a larger delocalization, the shift is maximized for S' excitons recombining with a polarization along the B family direction, while it is comparable to S excitons for those recombining along the A direction (Fig.~\ref{fig:fig4}\textbf{d}). This suggests non-trivial selection rules for the polarization of the resulting emitted photons that remain to be established. The spatial correlation discriminated by analyzing polarization is shown in SI. This demonstrates an emerging axial sensitivity, supporting the picture of an exciton recombining after two hops.

\vspace{3mm}
\section*{Discussion}
The observed shifts in energy with the in-plane strain of a few meV/\% point toward very localized species compared to Coulomb-dominated Wannier-Mott excitons in transition-metal dichalcogenides where the shift can be typically two orders of magnitude greater~\cite{Conley2013,Lloyd2016,Lopez2022}, approaching a behavior closer to quantum dots or defect-like emitters~\cite{Yeo2014,Kremer2014,Ovartchaiyapong2014,Wu2025}. This observation is consistent with the description of the excitation as a triplet-singlet transition within two wavefunctions centered on a single Ni site and spreading on neighboring S and P atoms~\cite{Hamad2024,Wang2024}.  
The excitons recombining in a magnetic configuration modified by the spins misalignment induced by the motion of the excitation through the lattice~\cite{Hamad2024,Liu1992} are accordingly less and less localized to the point that an axial dependency emerges. Raman spectroscopy, monitoring the energy evolution of the first magnon mode under stress, would be relevant to further discriminate the microscopic effect between the bare exciton and the dressed ones. The rules dictating the polarization of the photons emitted by the recombination of the exciton after two hops, seemingly favoring the secondary zigzag orientation under stress, will require further investigation. We note the absence of observable signature from the second zone-center antiferromangetic magnon, expected at 5\,meV~\cite{Jana2023, Na2024}.

The two nematic families seem robust in our experimental conditions (see SI). Compared to recent reports of nematicity in few-layer NiPS$_3$, we establish the coexistence of the two nematic domains within the sub-$\mu$m laser spot size which raises the question of their extension and condition of creation~\cite{Sun2024, Chen2024}. Nematic states may become easier to manipulate with very few layers, typically on a 5-layer membrane to maintain proper optical signatures. 
Strong anisotropic strain~\cite{Houmes2023} could be engineered to more accurately control magnon-dressed excitons. Future developments could involve alloys of NiPS$_3$~\cite{Lee2021,Kim2023a,Yan2023} to chemically tune the magnetic configuration and experience the sensitivity of the emission phenomena to impurities. 
Photoluminescence excitation spectroscopy combined with time-resolved PL could provide decisive insights into the exciton formation and recombination yields. 
Dynamical hybrid optomechanical coupling could be reached on these excitons to explore the tripartite coupling between phonons, magnons and excitons~\cite{Arcizet2011,Yeo2014,Barfuss2015,Lyons2023}.

\section*{Online Methods}
\subsection*{Sample preparation and characterization}
The electrodes in Ti(3\,nm)/Au(47\,nm) are deposited by e-beam evaporation after laser lithography of a UV-sensitive resist on top of a 500\,nm-thick SiO$_2$ substrate on a p-doped Si wafer. 5\,$\mu m$-diameter holes are made by reactive ion etching 
from the SiO$_2$ substrate. The etching depth ($464\pm14$\,nm) is measured by scanning electron microscopy.
The flakes are mechanically exfoliated from bulk crystals and dry-transferred on the substrate with a motorized transfer station in an Ar-filled glovebox. The thickness of the flakes are determined by a combination of atomic force microscopy, optical contrast and photoluminescence spectroscopy.
\subsection*{Optical setup}
The incident laser beam at 633\,nm is linearly-polarized, with typical optical power in the 100\,$\mu$W range and focused on the sample with a 0.65 NA cryogenic compatible microscope objective. The sample is thermo-regulated at 5.5\,K and its position by respect to the laser is locked with piezostages using the laser DC reflectance measured on a photodiode.
 
For optomechanical measurements (Fig.~\ref{fig:fig1}\textbf{d}), the incident photons are circularly polarized, through a polarizing beam splitter followed by a quarter-wave plate, such that the photons reflected by the sample can travel back and be reflected by the polarizing beam splitter, towards an avalanche photodiode. A network analyzer acquires the photodiode modulation at the electrostatic drive frequency. The static voltage V$_\mathrm{DC}$ is applied by a stabilized power supply, and mixed with the alternative voltage $V_\mathrm{AC}$ through a lab-made bias tee to limit cross-talks.  

The optical spectrometer includes a monochromator equipped with different gratings, in particular 600 and 1800 grooves/mm. The dispersed light is collected on a nitrogen-cooled CCD camera. For polarized optical spectroscopy, the incident light linear polarization is defined by a polarizer. A beam splitter placed just before the sample allows the collection of the scattered and emitted light, going then through an analyzer to discriminate its polarization. A half-wave plate is used before the spectrometer to  maintain a constant incident polarization on the monochromator gratings. For Raman measurements (Fig.~\ref{fig:fig3}\textbf{b}), an additional half-wave plate is placed right before the sample and rotated concomitantly with the analyzer angle to maintain a co-polarization configuration. The Raman features involving A$_g^2$ and B$_g^2$ phonons have maximum intensity in co-polarization and cross-polarization configuration, respectively (see SI). 

\subsection*{Acknowledgments}
We thank the STnano clean room staff, the SEM platform (MEB-Cro), M. Rastei, L. Engel, A. Boulard and R. Baehr for technical support, as well as Q. Fenoy, D. Voiry and C. Faugeras for fruitful discussions.
We acknowledge financial supports from the Agence Nationale de la Recherche (under grants VANDAMME ANR-21-CE09-0022, ATOEMS ANR-20-CE24-0010 and EXODUS ANR-23-QUAC-0004), 
from France 2030 government investment plan managed by the French ANR under grant reference PEPR SPIN – SPINMAT ANR-22-EXSP-0007, 
from the Interdisciplinary Thematic Institute QMat, as part of the ITI 2021-2028 program of the University of Strasbourg, CNRS and Inserm, by IdEx Unistra (ANR-10-IDEX-0002) and by SFRI STRAS'US (ANR-20-SFRI-0012) and EUR QMAT ANR-17-EURE-0024 under the framework of the French Investments for the Future Program, and by  
the Seed Money programme of Eucor – The European Campus. 

\bibliography{biblio_formate_url}
\end{document}